\documentstyle[aps,epsfig]{revtex}
\newcommand{\be}{\begin{equation}}
\newcommand{\ee}{\end{equation}}
\newcommand{\bea}{\begin{eqnarray}}
\newcommand{\eea}{\end{eqnarray}}
\newcommand{\bd}{\begin{displaymath}}
\newcommand{\ed}{\end{displaymath}}

\begin{document}
\bibliographystyle{physics}
\renewcommand{\thefootnote}{\fnsymbol{footnote}}

\begin{titlepage}

\begin{center}

{\large{\bf Excited heavy baryon masses in HQET QCD sum rules}}

\vspace{1.2cm}

C.-S. Huang$^a$, Ailin Zhang$^a$ and Shi-Lin Zhu$^{a,b}$

\vspace{0.8cm}
$^a$ Institute of Theoretical Physics, Academia Sinica, P.O. Box 2735, Beijing 100080, China\\
$^b$ Department of Physics, University of Connecticut,
Storrs, CT 06269 USA

\end{center}

\vspace{1.0cm}

\begin{abstract}

We construct the interpolating currents for the fourteen p-wave excited heavy baryons in 
the framework of heavy quark effective theory (HQET).
At the leading order in the $1/m_Q$ expansion these interpolating currents do not mix 
different states with the same $J^P$ with each other. We use QCD sum rule to calculate   
the masses of the p-wave excited J= $\frac{1}{2}$ spin doublet heavy baryons in the 
$m_Q \rightarrow \infty$ limit. We discuss some implications of our results.

\vskip 0.5 true cm
PACS Indices: 14.20.Lq, 14.20.Mr, 12.38.Lg

\end{abstract}

\vspace{2cm}
\vfill
\end{titlepage}

\section{Introduction}

In the past years important progress has been made in the field of heavy baryons.
Most of the ground state charmed baryons have been found experimentally while
among the bottomed baryons only $\Lambda_b$ is established \cite{pdg}. The lowest lying
orbitally excited states of $\Lambda_c$ have been observed by ARGUS \cite{argus}, 
E687 \cite{e687} and CLEO \cite{cleo} collaborations in the decay channel $\Lambda_c \pi\pi$.
The decay width of $\Lambda_{c1}$ with $J^P={1\over 2}^-$ is $3.6^{+2.0}_{-1.3}$ MeV \cite{pdg}.
For the state $\Lambda_{c1}$ with $J^P={3\over 2}^-$ only an upper limit of $1.9$ MeV
is set \cite{pdg}. Preliminary evidence of excited baryon with charm and strange quarks 
in the decay channel $\Xi^+ \pi \pi$ was reported by CLEO collaboration \cite{cleo1}.
Its width is less than 2.4 MeV.

Theoretically the heavy quark effective theory (HQET) \cite{hqet} provides a consistent framework
to study baryons containing one heavy quark in terms of $1/m_Q$ expansion with $m_Q$ the heavy
quark mass. Heavy baryon mass $m_B$ can be expanded as $m_B =m_Q +\bar \Lambda +{\cal
O}(1/m_Q)$ where $\bar \Lambda$ is the heavy baryon binding energy in the leading order.
Corrections at higher orders of the heavy quark expansion can be included consistently.
However in order to extract $\bar\Lambda$ we have to turn to other nonperturbative methods
such as lattice gauge theory, QCD sum rules (QSR)\cite{svz} etc for guidance. 
In this work
we will use the latter method to calculate the masses, or equivalently, the binding energies
 of the p-wave orbitally excited
heavy baryons in the leading order. The spectrum of ground state heavy baryons have been
studied with HQET using QSR in Refs.
\cite{g1,g2,g3}. Recently the binding energy of the lowest two states of orbitally excited charmed
baryons $\Lambda_{c1}, \Lambda_{c1}^*$ has been calculated \cite{zhu,liu}. 
However, the interpolating currents creating all the p-wave excited states
have not been given. 
In this work we construct the interpolating currents for the fourteen p-wave excited heavy baryons in 
the framework of heavy quark effective theory (HQET) and
 present a complete calculation of the spectrum of the p-wave excited spin 1/2 doublet heavy baryons at the leading order in the 1/$m_Q$ expansion.
Our calculation confirms earlier observation that the 
gluon condensates are of opposite sign to the leading perturbative term \cite{zhu}.

Excited heavy baryons are natural labs to test theoretical frameworks like HQET and explore 
the strong interaction dynamics which simplifies in the limit of $m_Q\to \infty$. Analyses of
various decays can also be used to extract some basic parameters of standard model. There 
are many theoretical papers in the recent years. The strong decays of the excited heavy baryons
were studied within the framework of heavy baryon chiral perturbation theory \cite{cho,dai}, 
various versions of quark model \cite{yan,korner,t-2,t-4,t-1,liu1,i-1}, and light cone QCD
sum rules \cite{zhu}. The electromagnetic decays were analyzed using a bound state picture in 
\cite{chow}, quark models in \cite{t-3,i-1} and light cone QCD sum rules \cite{zhu}.
Our construction of proper interpolating currents and calculation of leading order binding energy 
will faciliate the future study of pionic, electromagnetic and semileptonic decays of these excited
states within the framework of QSR. 

\section{Interpolating currents}

For a heavy baryon composed of a heavy quark Q and light freedom degrees, i.e., a light diquark 
system (qq), the spin-parity $j^P$ of the light diquark system is conserved in the $m_Q\rightarrow
\infty$ limit. Given $j^P$, one has a degenerate heavy baryon spin doublet with $J^P=(j\pm 1/2)^P
$. The heavy quark symmetry structure of heavy baryons is completely determined by the spin-
parity $j^P$ of the light diquark system.

A systematic classification of heavy baryon states by constituent approach has been presented~\cite
{hklt}. For p-wave excited heavy baryons, there are two independent orbital angular momenta
$l_k$ and $l_K$ corresponding to the two independent momenta k and K which we take to be the
relative momenta k=$\frac{1}{2}(p_1-p_2)$ and K=$\frac{1}{2}(p_1+p_2-2p_3)$ where $p_1$ and $p_2$
are the light quark momenta and $p_3$ the heavy quark momentum. The choice (k, K) basis has
two advantages. First, with such a choice the physical meaning of $l_k$ and $l_K$ is transparent:
the k-orbital angular momentum $l_k$ describes relative motion of the two light quarks, and the
K-orbital angular momentum $l_K$ describes orbital motion of the center of mass of the two light
quarks relative to the heavy quark. Second, the (k, K) basis allows one to classify the
diquark states in terms of SU(2$N_f$)$\otimes$O(3) representations~\cite{hklt} where $N_f$
is the number of light flavors. We assume $N_f$=2 in this paper and generization to the
$N_f$=3 case is straightforward.

\vspace{2cm}
\begin{table}
  \begin{tabular}{lllllll}
  Symmetric property&$j^P$&$J^P$&State\\
 \hline\hline
             &             &$\frac{1}{2}^-$& \\
 $[q_1 q_2]$:&$1^- \begin{array}{ll}
                     \nearrow \\ \searrow
                    \end{array}
                     $&  &  $\{\Lambda_{QK1}\}$\\
             &               &$\frac{3}{2}^-$&\\
\hline

$\{ q_1 q_2 \}$:&$0^-\rightarrow$&$\frac{1}{2}^-$& $\Sigma_{QK0}$\\
               &             &$\frac{1}{2}^-$& \\
                &$1^- \begin{array}{ll}
                      \nearrow \\ \searrow
                      \end{array}$&  &  $\{\Sigma_{QK1}\}$\\
                &               &$\frac{3}{2}^-$&\\
             &             &$\frac{3}{2}^-$& \\
                &$2^- \begin{array}{ll}
                \nearrow \\ \searrow \end{array}$&  &  $\{\Sigma_{QK2}\}$\\
                       &               &$\frac{5}{2}^-$& .\\ 
  \hline\hline 
\end{tabular}
  \caption{The p-wave heavy baryon states, with $l_k$=0 and $l_K$=1.}
\label{baryons-1}
\vspace{5mm}
\end{table}
According to the analysis in ref.~\cite{hklt}, the p-wave excitation with $l_k$=0 and $l_K$=1
belongs to the representation 10 $\otimes 3_K$ of SU(4)$\otimes$ O(3). Under the $SU(2)_{spin}
\otimes SU(2)_{flavor}$ the 10 decomposes into $1\otimes 1$ and $3\otimes 3$. The spin 0 and 1
pieces of the 10 couple with $l_K$=1 to give $j^P= 1^-$ state with flavor anti-symmetric (
$\Lambda$-type) and $j^P=0^-, 1^-, 2^-$ states with flavor symmetric ($\Sigma$-type), respectively.
The total angular momentum j of a diquark with definite parity P couples with the spin of the
heavy quark finally to give the p-wave heavy baryon with spin-parity $J^P$ states which are listed
in Table 1.
Similar analyses apply to the p-wave excitations with $l_k$=1 and $l_K$=0
and results are listed in Table 2.

Thus we have total fourteen p-wave heavy baryon states with negative parity of which seven
states are $\Lambda$-type and the others $\Sigma$-type, corresponding to $[q_1q_2]=q^T\tau_A
q$ with $q^T=(q^T_1, q^T_2)$ and $\{q_1 q_2\}=q^T\tau_i q$ (i=1,2,3) (for the definition of $\tau_B$
, B=1,2,3,A, see below), respectively. We denote them by $B_{Qpj}$ with
B=$\Lambda, \Sigma$, Q=b,c, p=k,K and j=0,1,2.

An important step to carry out QCD sum rule analysis of Green functions is to construct 
appropriate interpolating currents which create corresponding heavy baryon states. For the
above fourteen p-wave states, we propose to use 
\be
\bar{j}_{\rho_1...\rho_J}^{B,p,j,J}\equiv j_{\rho_1...\rho_J}^{\dagger~ B,p,j,J} \gamma^0
\ee
as such interpolating currents. Here\\ 
\begin{table}
  \begin{tabular}{lllllll}
  Symmetric property&$j^P$&$J^P$&State\\
 \hline\hline
              $[q_1 q_2]$:&$0^- \rightarrow$&$\frac{1}{2}^-$& $\Lambda_{QK0}$\\
              &             &$\frac{1}{2}^-$& \\
         &$1^-      \begin{array}{ll}
                     \nearrow \\ \searrow

                    \end{array}
                     $&  &  $\{\Lambda_{QK1}\}$\\
                   &               &$\frac{3}{2}^-$&\\
                 &             &$\frac{3}{2}^-$& \\
                &$2^- \begin{array}{ll}
                \nearrow \\ \searrow \end{array}$&  &  $\{\Lambda_{QK2}\}$\\
                       &               &$\frac{5}{2}^-$& .\\ 
  \hline
$\{ q_1 q_2 \}$:
               &             &$\frac{1}{2}^-$& \\
                &$1^- \begin{array}{ll}

                      \nearrow  \\ \searrow

                      \end{array}$&  &  $\{\Sigma_{QK1}\}$\\
                &               &$\frac{3}{2}^-$&\\
            \hline\hline 
\end{tabular}
  \caption{The p-wave heavy baryon states, with $l_k$=1 and $l_K$=0.}
\label{baryons-2}
\end{table}
\be
j_{\rho_1...\rho_J}^{B,p,j,J}(x)=\epsilon^{abc} ( q^T_a(x)\tau_B (a +
b \rlap/ v )\phi^{\mu_1...\mu_j}_p C
q_b(x)) \Gamma^J_{\mu_1...\mu_j;\rho_1...\rho_J} h_{v,c}(x),
\ee
where
\be
\phi^{\mu_1...\mu_j}_p=\phi^{\mu_1...\mu_j;\nu}(v) D_{\nu}^p,\hspace{1cm}
 D_{\nu}^p= \left \{
\begin{array}{ll}
\displaystyle{\frac{{\partial}}{\partial x^{\nu}}-ig A_{\nu}(x)} & \mbox{for p=k}\\
\displaystyle{\frac{{\partial}}{\partial y^{\nu}}-ig A_{\nu}(y)} & \mbox{for p=K} 
\end{array} \right. 
\ee
with $x$ and $y$ the relative coordinates between light quarks and between
the center of mass
of light quark system and the heavy quark respectively.
$D_{\mu}(x)$ operates on a light quark field q and $D_{\mu}(y)$ on 
the effective heavy quark field $h_v$ (see Table 3 for details).
\be
\Gamma^J_{\mu_1...\mu_j;\rho_1...\rho_J} = \left \{
\begin{array}{ll}
\Gamma^{j+1/2}_{\mu_1...\mu_j;\rho_1...\rho_j} &
\mbox{for J=j+$\frac{1}{2}$} \\
\Gamma^{j-1/2}_{\mu_1...\mu_j;\rho_1...\rho_{j-1}} & \mbox{for J=j-$\frac{1}{2}$}
\end{array} \right.
\ee
consists of Dirac $\gamma$ matrices and the covariant derivative, and $\tau_B$ (B=1,2,3,A) are
defined by
\be
\tau_1  =  \frac{1}{\sqrt{2}}\left(
\begin{array}{lr}
 0 & 1 \\
 1 & 0
 \end{array} \right), ~~~~~~~~~
\tau_2=\left( \begin{array}{lr} 1 & 0 \\ 0 & 0 \end{array} \right),
\tau_3 =\left( \begin{array}{lr} 0 & 0 \\ 0 & 1 \end{array} \right),~~~~~~~~~
\tau_A=\frac{1}{\sqrt{2}}\left( \begin{array}{lr} 0 & 1 \\ -1 & 0 \end{array} \right).
\ee
 \vspace{5mm}
\begin{table}
  \begin{tabular}{lllllll}
 States & $\phi^{\mu_1...\mu_j}_p$ &  $j^P$ & $\Gamma^J_{\mu_1...\mu_j;\rho_1...\rho_J}$
&  $J^P$\\
 \hline\hline
p=k, i.e., $l_k=1, l_K=0$ \\
   $\{\Sigma_{Qk1}\}$ & $\gamma_5 D_t^{\mu_1}(x)$ & $1^-$ &
$    \begin{array}{ll}$ 
  $\gamma_{t \mu_1}\gamma_5  \\  \Gamma_{\mu_1\rho_1}$
 $ \end{array} $  & 
  $  \begin{array}{ll}$
  $\frac{1}{2}^-  \\  \frac{3}{2}^-$
$ \end{array}  $         \\
$\{\Lambda_{Qk0}\}$ & $\gamma_t^{\mu} D_{t\mu}(x)$ & $0^-$ & I & $\frac{1}{2}^-$ \\
$\{\Lambda_{Qk1}\}$ &
$\epsilon^{\mu_1\nu\sigma\rho}\gamma_{t\nu}D_t^{\sigma}(x)v^{\rho}$ & $1^-$ &
$ \begin{array}{ll} $
$\gamma_{t \mu_1}\gamma_5  \\  \Gamma_{\mu_1\rho_1}$
$ \end{array} $ &
$ \begin{array}{ll} $
$ \frac{1}{2}^-  \\  \frac{3}{2}^-$
$\end{array}$           \\
$\{\Lambda_{Qk2}\}$ & $\{\gamma_t^{\mu_1} D_t^{\mu_2}(x)\}_0$ & $2^-$ &
$\begin{array}{ll}$
$\gamma_5\gamma_{t\{\mu_1} \Gamma_{\mu_2\}\rho_1}  \\  \Gamma_{\rho_1\{\mu_1}
\Gamma_{\mu_2\}_0\rho_2}$
$\end{array} $  &
$\begin{array}{ll} $
$\frac{3}{2}^-  \\  \frac{5}{2}^-$
$\end{array}      $     \\
\hline
p=K, i.e., $l_k=0, l_K=1$ \\
   $\{\Lambda_{QK1}\}$ & $\gamma_5 D_t^{\mu_1}(y)$ & $1^-$ &
$    \begin{array}{ll}$ 
  $\gamma_{t\mu_1}\gamma_5  \\  \Gamma_{\mu_1\rho_1}$
 $ \end{array} $  & 
  $  \begin{array}{ll}$
  $\frac{1}{2}^-  \\  \frac{3}{2}^-$
$ \end{array}  $         \\
$\{\Sigma_{QK0}\}$ & $\gamma_t^{\mu} D_{t\mu}(y)$ & $0^-$ & I & $\frac{1}{2}^-$ \\
$\{\Sigma_{QK1}\}$ &
$\epsilon^{\mu_1\nu\sigma\rho}\gamma_{t\nu}D_t^{\sigma}(y)v^{\rho}$ & $1^-$ &
$ \begin{array}{ll} $
$\gamma_{t \mu_1}\gamma_5  \\  \Gamma_{\mu_1\rho_1}$
$ \end{array} $ &
$ \begin{array}{ll} $
$ \frac{1}{2}^-  \\  \frac{3}{2}^-$
$\end{array}$           \\
$\{\Sigma_{QK2}\}$ & $\{\gamma_t^{\mu_1} D_t^{\mu_2}(y)\}_0$ & $2^-$ &
$\begin{array}{ll}$
$\gamma_5\gamma_{t\{\mu_1} \Gamma_{\mu_2\}\rho_1}  \\  \Gamma_{\rho_1\{\mu_1}
\Gamma_{\mu_2\}_0\rho_2}$
$\end{array} $  &
$\begin{array}{ll} $
$\frac{3}{2}^-  \\  \frac{5}{2}^-$
$\end{array}      $     \\
\hline\hline
\end{tabular}
  \caption{The explicit expressions of $\phi^{\mu_1...\mu_j}_p$ and $\Gamma^J_{\mu_1...\mu_j;\rho_1...
\rho_J}$ for the p-wave states. Here I is the 4$\times$4 unit matrix and $\Gamma_{\mu\nu}=
-\frac{1}{3}(g_{t\mu\nu}+\gamma_{t\nu}\gamma_{t\mu})$ with 
$\gamma_t^\mu=\gamma^\mu-v^\mu\rlap/\!v$. 
$D_{\mu}(x)$ operates on
a light quark field q and $D_{\mu}(y)$ the effective heavy quark field
$h_v$. }
\label{baryons-3}
\end{table}
The matrices $\tau_B$ describe the flavor structures of the diquark inside a heavy baryon and
satisfy
\be
tr (\tau_B \tau_{B^{'}}^{\dagger}) = \delta_{BB^{'}},~~~~~~B,B^{'}=1,2,3,A.
\ee
The explicit expressions of $\phi^{\mu_1...\mu_j}_p$ and $\Gamma^J_{\mu_1...\mu_j;\rho_1...
\rho_J}$ for the p-wave states are listed in Table 3.
Let $\mid B,p,j,J >$ be a p-wave excited heavy baryon state with quantum numbers j, J and
definite flavor type B and orbit angular momentum type p in the $m_Q\rightarrow \infty$
limit. Because we limit ourselves only to p-wave excited states which all have
negative parity in this work, we omit the parity index. We have
\be
<0\mid j_{\rho_1...\rho_J}^{B,p,j,J}(0)\mid B^{'}, p^{'}, j^{'}, J^{'}>=
\delta_{BB^{'}}\delta_{pp^{'}}
\delta_{jj^{'}}\delta_{JJ^{'}} f_{j,J} u_{\rho_1...\rho_J}.
\ee
In order to verify eq.(7) an easy way is to choose the gauge $A_{\mu}(x_0)=0$ without loss of
generality. In the gauge $D_{\mu}$(x) is reduced to $\frac{\partial}{\partial x^{\mu}}$ and the Bethe-Salpeter(BS)
wave function of a p-wave excited heavy baryon is defined by
\bea
{\large \chi}^{\alpha\beta\gamma;ii^{\prime}}_{B,p,j,J}(x_1,x_2,x_3) & = & <0 \mid
T(q^i_{\alpha}(x_1)q^{i^{\prime}}_{\beta}(x_2)Q_{\gamma}(x_3)) \mid B,p,j,J > \nonumber   \\
& = & e^{-iP\cdot Y}{\large \chi}^{\alpha\beta\gamma;ii^{\prime}}_{B,p,j,J}(x,y)  \\
Y&=&x_1+x_2+x_3,~~~~~~x=(x_1-x_2),~~~~~~~y=(x_1+x_2-2x_3)/3 \nonumber
\eea
with $\alpha,\beta,\gamma$ being Dirac indices and P the momentum of baryon.
Here and hereafter colour indices are suppressed for the sake of simplicity. As shown in ref.
\cite{hklt}, in the $m_Q\rightarrow \infty$ limit, the BS wave function can be written as
\be
{\large \chi}^{\alpha\beta\gamma;ii^{\prime}}_{B,p,j,J}(x,y)=\tau^{\dagger ii^{\prime}}_B C
\bar{\phi}^{\mu_1...\mu_j}_{p\alpha
\beta}(v,x,y) (1 + \rlap/ v ) f(x,y)\Psi^{J,\gamma}_{\mu_1...\mu_j},
\ee
where f(x,y)=$f(x_t^2,y_t^2,x_l,y_l)$ is a Lorentz scalar function of $z_t^2$ and $z_l$ (z=x,y),
$z_l=z\cdot v$, $z_t=z-z_l v$, and $\Psi^{J}_{\mu_1,...,\mu_j}$ is given by
\bea
\Psi^{\frac{1}{2}}&=&u,  \nonumber \\
\Psi^{\frac{1}{2}}_{\mu_1}&=&\gamma_{t\mu_1}\gamma_5 u,  \nonumber \\
\Psi^{\frac{3}{2}}_{\mu_1}&=&u_{\mu_1},     \nonumber \\
\Psi^{\frac{3}{2}}_{\mu_1\mu_2}&=&\gamma_5 \gamma_{t\{\mu_1}
u_{\mu_2\}_0},  \nonumber \\
\Psi^{\frac{5}{2}}_{\mu_1\mu_2}&=&u_{\mu_1\mu_2},
\eea
where the notation $\{\mu_1\mu_2\}_0$ implies symmetrization and
tracelessness, $\gamma_t^\mu=\gamma^\mu-v^\mu\rlap/\!v$ with $v^\mu$ is
the velocity of the heavy baryon, u and 
$u_{\mu_1...\mu_j}$ are the usual Dirac spinor and the Rarita-Schwinger spinor 
respectively. The latter satisfies
\be
\gamma^{\mu_i} u_{\mu_1...\mu_i...\mu_j}=0,~~~~~ v^{\mu_i} u_{\mu_1...\mu_i...\mu_j}=0
\ee
and are the symmetric and traceless.
Therefore, from eqs. (2) and (9), we have
\bea
<0 \mid j_{\rho_1...\rho_J;\gamma}^{B,p,j,J}(x_0) \mid B^{'}, p^{'}, j^{'},
 J^{'}>&=&\tau_B^{ii^{\prime}} (a + b \rlap/ v )
\phi^{\mu_1...\mu_j}_{p~~\alpha\beta} C (\Gamma^J_{\mu_1...\mu_j;\rho_1...\rho_J})^{\gamma\lambda}
 \nonumber \\ 
& &{\large \chi}_{B^{'}p^{'},j^{'},J^{'}}^{\beta\alpha\lambda;i^{\prime}i}(x_0
+\frac{3y}{2}+\frac{x}{2}, x_0+\frac{3y}{2}-\frac{x}{2}, x_0)\mid_{x=y=0}  \nonumber \\
 &=& tr(\tau_B\tau_{B^{'}}^{\dagger})~~ tr(\phi^{\mu_1...\mu_j}_p\bar{\phi}^
{\nu_1...\nu_j}_{p^{'}})~~(a+b)~~ f  \nonumber \\
& &(\Gamma^J_{\mu_1...\mu_j;\rho_1...\rho_J}\Psi^{J^{'}}_{\nu_1...\nu_{j^{'}}})^
{\gamma}\mid_{x=y=0}~~e^{-i\bar{\Lambda}v\cdot x_0}.
\eea
Using
\be
\frac{\partial^n f(x_t^2,y_t^2,x_l,y_l)}{\partial z_t^{\mu_1}...
\partial z_t^{\mu_n}}\mid_{x_t=y_t=0}=0~~~~~~for~~ n=odd,
\ee
and
\be
\frac{\partial^2  f(x_t^2,y_t^2,x_l,y_l)}{\partial z^{\mu}_t\partial
 z^{'\nu}_t}\mid_{x_t=y_t=0}
=\delta_{z_t z_t^{'}} g^{\mu\nu} 2 \frac{\partial^2 f}{\partial z^2_t}
\mid_{z_t=0}
\ee
with z, $z^{'}$=x, y,
one has 
\bea
tr(\phi^{\mu_1...\mu_j}_p\bar{\phi}^{\nu_1...\nu_{j^{'}}}_{p^{'}})f &=&
tr(\phi^{\mu_1...\mu_j;\mu}\bar{\phi}^{\nu_1...\nu_{j^{'}};\mu})\delta_
{pp^{'}} 2 \frac{\partial^2 f}{\partial z^2_t}\mid_{z_t=0}  \nonumber \\
  &=&\delta_{jj^{'}}\delta_{pp^{'}} G^{\mu_1...\mu_j;\nu_1...\nu_j}
2\frac{\partial^2 f}{\partial z^2_t}\mid_{z_t=0},
\eea
where 
\bea
G&=&1~~~~~~~~~for~~~~ j=0, \nonumber \\
G^{\mu\nu}&=&-g^{\mu\nu}_t\equiv g^{\mu\nu}-v^{\mu}v^{\nu}~~~~~~~~for~~~~ j=1,
 \nonumber \\
G^{\mu_1\mu_2;\nu_1\nu_2}&=&\frac{1}{2}(g_t^{\mu_1\nu_1}g_t^{\mu_2\nu_2}+g_t^{\mu_1\nu_2}
g_t^{\mu_2\nu_1}-\frac{2}{3}g_t^{\mu_1\mu_2}g_t^{\nu_1\nu_2})~~~for~~~~ j=2.
\eea
By using eqs.(10, 16) and the explicit expressions for $\Gamma^J_{\mu_1...\mu_j;\rho_1...\rho_J}$
 given in Table 3, it is now straightforward to derive
\be
G^{\mu_1...\mu_j;\nu_1...\nu_j}\Gamma^J_{\mu_1...\mu_j;\rho_1...\rho_J}
\Psi^{J^{'}}_{\nu_1...\nu_j}=\delta_{JJ^{'}} N_{J,j} u_{\rho_1...\rho_j},
\ee
where $N_{J,j}$ is a number dependent of J and j. Combining eqs.(6),(12),(15),and (17), one  
arrives at eq.(7).
If we normalize the current $j^{B,p,j,J}_{\rho_1...\rho_j}$ appropriately, the baryonic
"decay constant" $f_{j,J}$ can be written as $f_{j,J}$=$N_J f_j$ 
\vspace{5mm}
\begin{table}
  \begin{tabular}{lllllll}
 States & $\phi^{\mu_1...\mu_j}_p$ &  $j^P$ & $\Gamma^J_{\mu_1...\mu_j;\rho_1...\rho_J}$
&  $J^P$\\
 \hline\hline
p=K, i.e., $l_k=0, l_K=1$ \\
   $\{\Lambda_{QK1}\}$ & $\gamma_5 \gamma^{\mu_1}$ & $1^-$ &
$    \begin{array}{ll}$ 
  $\gamma_t^{\mu_1}\gamma_5 \,\,  \rlap/\! D_t \\  \Gamma_{\mu_1\rho_1}\,\,\rlap/\! D_t$
 $ \end{array} $  & 
  $  \begin{array}{ll}$
  $\frac{1}{2}^-  \\  \frac{3}{2}^-$
$ \end{array}  $         \\
$\{\Sigma_{QK0}\}$ & $\gamma_t^{\mu}$ & $0^-$ & $\gamma_{t\mu}\,\,\rlap/\! D_t$ & $\frac{1}{2}^-$ \\
$\{\Sigma_{QK1}\}$ &
$\epsilon^{\mu_1\nu\sigma\rho}\gamma_{t\nu}v^{\rho}$ & $1^-$ &
$ \begin{array}{ll} $
$\gamma_t^{\mu_1}\gamma_5 \gamma^{\sigma}_t\,\,\rlap/\! D_t \\  \Gamma_{\mu_1\rho_1} \gamma^
{\sigma}_t \,\,\rlap/\! D_t$
$ \end{array} $ &

$ \begin{array}{ll} $
$ \frac{1}{2}^-  \\  \frac{3}{2}^-$
$\end{array}$           \\
\hline\hline
\end{tabular}
  \caption{Some examples of other possible interpolating currents. }
\label{table-4}
\end{table}
where $f_j$ only depends on
the diquark spin j and has the same value for the two states in the same doublet.\\
Similarly, using the same method in ref.~\cite{dhhl}, we can verify
\be
<0 \mid T(j^{B,p,j,J}_{\rho_1...\rho_j}(x)\bar{j}^{B^{'},p^{'},j^{'},
J^{'}}_{\rho_1^{'}...\rho_j^{'}}(0))\mid 0>
\propto \delta_{BB^{'}}\delta_{pp^{'}}\delta_{jj^{'}}\delta_{JJ^{'}} .
\ee
Eqs.(7) and (18) imply that two currents with different B,p,j,J never mix in the $m_Q\rightarrow
\infty$ limit. Therefore, eq.(1) are the appropriate interpolating currents for p-wave excited
heavy baryons. We would like to point out that for higher (e.g., D-wave) excited states to
construct interpolating currents with such properties are possible but not easy due to the presence of
many different orbital states with given total orbital angular momentum of diquark.

It is well-known that the choice of interpolating currents is not unique. For example, the
following interpolating currents in Table \ref{table-4} have also the above mentioned properties.

\section{The mass sum rules}

Since the sum rule for $\Lambda_{QK1}$ is already presented in \cite{zhu,liu}, we 
consider the sum rules for the states $\Sigma_{QK0}, \Sigma_{QK1},
\Sigma_{Qk1}, \Lambda_{Qk0}, \Lambda_{Qk1}$. To be specific, 
we use the interpolating currents listed in tables 3 and 4 in our calculation :
\begin{equation}\label{q1}
j_{\Sigma_{QK0}}(x)=\epsilon^{abc}[q^T_a\tau_B C\gamma^\mu_tq_b]
\gamma_{t \mu}\stackrel{\rightarrow}{\rlap/\! D_t}h_c(x)
\end{equation}
\begin{equation}
j_{\Sigma_{QK1}}(x)=\epsilon^{abc}\epsilon_{\mu\nu\rho\sigma}v^\rho[q^T_a\tau_B   C\gamma^\nu_tq_b]
\gamma^\mu_t\gamma^\sigma_t\stackrel{\rightarrow}{\rlap/\! D_t}\gamma_5h_c(x),
\end{equation}
\begin{equation}
j_{\Sigma_{Qk1}}(x)=\epsilon^{abc}[q^T_a\tau_B  C\gamma_5\stackrel{\rightarrow}{D_{t \mu}}q_b]
\gamma^\mu_t\gamma_5h_c(x),
\end{equation}
\begin{equation}
j_{\Lambda_{Qk0}}(x)=\epsilon^{abc}[q^T_a\tau_B  C\gamma^\mu_t
\stackrel{\rightarrow}{D_{t \mu}}q_b]h_c(x)
\end{equation}
\begin{equation}\label{q2}
j_{\Lambda_{Qk1}}(x)=\epsilon^{abc}\epsilon_{\mu\nu\rho\sigma}v^\rho[q^T_a\tau_B  C\gamma^\nu_t
\stackrel{\rightarrow}{D^\sigma_t}q_b] 
\gamma^\mu_t\gamma_5h_c(x),
\end{equation}
where $a$, $b$, $c$ is the color index, $T$ denotes the transpose, $C$ is the charge conjugate
matrix, $\gamma_t^\mu$ is defined as above. Considering
the result in the ref.~\cite{liu} that the best stability of the sum rule for the current with a derivative is obtained for a=1 and b=0, we have taken a=1 and b=0 in eqs. (19-23) for the sake of simplicity.
We also need
\begin{equation}
\bar j_{\Sigma_{QK0}}(x)=\epsilon^{abc}\bar
h_c\stackrel{\leftarrow}{\rlap/\! D_t}\gamma_{t\mu}[\bar
q_b\gamma^\mu_t\tau_B^\dagger C \bar q^T_a]
\end{equation}
\begin{equation}
\bar j_{\Sigma_{QK1}}(x)=
-\epsilon^{abc}\epsilon_{\mu\nu\rho\sigma}v^\rho
\bar
h_c\gamma_5\stackrel{\leftarrow}{\rlap/\! D_t}\gamma^\sigma_t\gamma^\mu_t
[\bar q_b\gamma^\nu_t\tau_B^\dagger C \bar q^T_a],
\end{equation}
\begin{equation}
\bar j_{\Sigma_{Qk1}}(x)=-\epsilon^{abc}\bar h_c\gamma^\mu_t
\gamma_5[\bar q_b\stackrel{\leftarrow}{D_{t\mu}}\gamma_5\tau_B^\dagger C
\bar q^T_a],
\end{equation}
\begin{equation}
\bar j_{\Lambda_{Qk0}}(x)=\epsilon^{abc}\bar
h_c[\bar q_b\stackrel{\leftarrow}{D_{t\mu}}\gamma^\mu_t\tau_B^\dagger C \bar q^T_a]
\end{equation}
\begin{equation}
\bar j_{\Lambda_{Qk1}}(x)=
\epsilon^{abc}\epsilon_{\mu\nu\rho\sigma}v^\rho
\bar h_c\gamma^\mu_t\gamma_5
[\bar q_b\stackrel{\leftarrow}{D^\sigma_t}\gamma^\nu_t\tau_B^\dagger C \bar q^T_a],
\end{equation}
The overlap amplitudes of the interpolating currents with the heavy baryons
are defined as follows.
\begin{equation}
\label{lap1}
\langle 0|j_{\Sigma_{QK1}} |\Sigma_{QK1}\rangle =f_{\Sigma_{QK1}}u_{\Sigma_{QK1}} \, .
\end{equation}
Similar definitions hold for other states.

In order to extract the binding energy of the p-wave heavy baryons at the
leading order in the $1/m_Q$ expansion, we consider the correlation function
\begin{equation}
{\it i}\int e^{{\it i}qx}\langle0|T\{j_i(x),\bar j_i(0)\}|0\rangle dx
={1+\rlap/v\over 2} \Pi_i(\omega),
\end{equation}
with $\omega=q\cdot v$.

The dispersion relation for $\Pi (\omega )$ reads
\begin{equation}\label{dip-1}
\Pi ( \omega )={1\over \pi}\int {\mbox{Im} \Pi (s)\over s- \omega  -i\epsilon }ds\;,
\end{equation}
where $\mbox{Im} \Pi (s) $ is the spectral density in the limit $m_Q \to \infty$.

At the phenomenological side we use
\begin{equation}
\mbox{Im} \Pi (s )= f^2 \delta (s-\Lambda)
+\mbox{Im} \Pi^{\mbox{Pert}} (s)  \theta (s-\omega_c ) \;,
\end{equation}
where we approximate the continuum or more higher states contribution above 
$m_Q+\omega_c$ with the perturbative contribution at the quark gluon level.
We invoke Borel transformation with the variable $\omega$ to
(\ref{dip-1}) to suppress the continuum contribution further. 
Finally we have
\begin{equation}
\label{mass-1}
f^2e^{-{\Lambda\over T}}
=\int_0^{\omega_c} \mbox{Im} \Pi (s) e^{-{s\over T}}ds\;.
\end{equation}

At the quark level the spectral density reads
\begin{equation}\label{s1}
\mbox{Im}\Pi_{\Sigma_{QK0}} (s)={11s^7 \over 140\pi^3}
-{\langle\alpha_sG^2\rangle s^3\over 96\pi^2},
\end{equation}
\begin{equation}
\mbox{Im}\Pi_{\Sigma_{QK1}}(s)={11s^7 \over 35\pi^3}
-{\langle\alpha_sG^2\rangle s^3\over 24\pi^2},
\end{equation}
\begin{equation}
\mbox{Im}\Pi_{\Sigma_{Qk1}}(s)={s^7 \over 112\pi^3}
+{3\langle\alpha_s G^2\rangle s^3\over 128\pi^2}
-{\langle\alpha_s G^2\rangle s^3\over 32\pi^2},
\end{equation}  
\begin{equation}
\mbox{Im}\Pi_{\Lambda_{Qk0}}(s)={\tau^7 \over
560\pi^3}-{7\langle\alpha_sG^2\rangle\tau^3\over  384\pi^2},
\end{equation}  
\begin{equation}\label{s2}
\mbox{Im}\Pi_{\Lambda_{Qk1}}(s)={3 s^7 \over 140\pi^3}
+{\langle\alpha_sG^2\rangle s^3\over 96\pi^2}
-{\langle\alpha_sG^2\rangle s^3\over 32\pi^2},
\end{equation}
with $\langle {\alpha_s\over \pi}G^2\rangle =0.012$GeV$^4$\cite{ruckl},
where the first term of the
gluon condensate in each spectral density results from the diagram with
one gluon cut attached on each light quark propagator, while the second
piece  arises from the diagrams with one gluon cut attached on the vertex.
An interesting feature of (\ref{s1})-(\ref{s2}) is that the gluon condensate is of
the opposite sign as the leading perturbative term, in contrast with the
ground state baryon mass sum rules. This may be interpreted as some kind of
gluon excitation since we are considering p-wave baryons. In the present case
the gluon in the covariant derivative also contributes to various condensates.

To obtain the numerical results for the leading order binding energy
$\Lambda$, the following formulae from the dispersion relation and
quark-hardon duality is used
\begin{eqnarray}
\Lambda={{1\over \pi}\int_{0}^{\omega_c} s e^{-s/\tau}
\mbox{Im}\Pi^{\mbox{pert}}(s) ds
-{d \over d 1/\tau}\hat{B}^\omega_\tau \Pi^{\mbox{cond}}(\omega) \over 
{1\over \pi}\int_{0}^{\omega_c} e^{-s/\tau} \mbox{Im}\Pi^{\mbox{pert}}(s)
ds 
+\hat{B}^\omega_\tau \Pi^{\mbox{cond}}(\omega) },
\end{eqnarray}
where the operator ${\hat B}$ denotes Borel transformation and $\Pi^{\mbox{cond}}(\omega)$
denotes the condensate contribution to $\Pi$ ($\omega$) in (\ref{dip-1}).

With the suitable choice of $\tau$ and $\omega$, the heavy baryon binding
energy $\Lambda$ is determined. All the numerical results are presented in
Fig. 1. It is easy to find that the sum rules for states
$\Sigma_{QK0}$, $\Sigma_{QK1}$, $\Sigma_{Qk1}$, and $\Lambda_{Qk1}$ are
stable in the variable region 0.2 GeV to 0.8 GeV. To find the most
suitable threshold $\omega_c$, we have tried several
choices: $\omega_c=1.4, 1.6$
and 1.8 GeV etc for each current. The final results are given at
$\omega=1.6$ GeV. In these sum rules 
 the gluon condensate contributions are $10-30\%$ of the perturbative term 
in the Borel variable region from 0.3 GeV to 0.7 GeV.
The lowest lying resonances dominate.  As a
by-product, the decay constants of these states are obtained. All the
obtained binding energy $\Lambda_i$ and the decay constant $f_i$ are
collected in Table \ref{table5}, where the variation of $f_i$ comes from
variation of Borel variable $\tau$.
 
Finally, some words on the current $j_{\Lambda_{Qk0}}$ should be said. In contrast with the numerical results for the currents above,
gluon condensate dominate the sum rule for the current $j_{\Lambda_{Qk0}}$  in the region with Borel variable
larger than zero, so we do not give its numerical results here. 
\begin{table}[b]
\caption{Binding energy of p-wave states.}
\vspace{1cm}
\label{table5}
\begin{center}
\begin{tabular}{c c c c c c}
-&$\Sigma_{QK0}$  &  $\Sigma_{QK1}$ & $\Sigma_{Qk1}$ & $\Lambda_{Qk1}$ \\
\hline
$\Lambda_i$ & 1.35 GeV & 1.32 GeV & 1.30 GeV & 1.28 GeV &\\  
$f_i$& $0.0339-0.0354 $GeV$^4$&$0.0669-0.0671 $GeV$^4$&$0.0188-0.0190 
$GeV$^4$&$0.0268-0.0286 $GeV$^4$&
\end{tabular}   
\end{center}
\end{table}   

With the renormalization-group and scheme
invariant pole mass for the heavy quarks
$m_c=(1.3-1.4)$ GeV and $m_b=(4.6-4.8) $GeV,
we obtain the spectrum of the p-wave excited spin 1/2 doublet baryon states.

\section{Discussions}

In summary we have constructed the proper interpolating currents suitable for
QCD sum rule
approach for the orbitally  excited heavy baryons in the framework of heavy quark 
effective theory. These currents are orthogonal to each other in the leading order 
of HQET and explore different excited states (some of them have the same quantum numbers).
The mixing of these currents for the same quantum numbers occur at the order of $1/m_Q$. 
We want to emphasize that only in the leading order of HQET can we construct 
interpolating currents with such good properties.
We obtain the leading order mass sum rules for the orbitally excited
states with $J\le
{3\over 2}$ in HQET. We find the gluon condensates
have opposite sign to the leading
perturbative terms which may indicate the orbital excitation of the gluon fields inside the
excited heavy baryons. All sum rules are stable with reasonable variations of the Borel
parameter and the continuum threshold. We extract the leading order binding energy and the 
overlapping amplitudes, which may be used to analyze the semileptonic decays of
these states. The spectrum and level spacing from our approach is consistent 
with that from quark model prediction \cite{korner}. Typically $\Lambda_{QK0}$ lies
300 MeV higher than $\Lambda_Q$ while all other excited states with $J\le {3\over 2}$
lie approximately 500 MeV higher than $\Lambda_Q$. However the $1/m_Q$ correction 
may be significant for the excited charmed baryons. Note that the magnitudes of the binding energy
we obtained are very close  to the charm quark pole mass. Due to the mixing of these interpolating
currents  the calculation of $1/m_Q$ correction is difficult and lies out of the scope of 
present paper.

\section*{Acknowledgment}

We would like to thank Y.-B. Dai for discussions. 
 C.-S. Huang and S.-L. Zhu were partly supported by National Natural Science 
Foundation of China.

{\bf Figure Captions}

\begin{center}
{\sf FIG 1.} {(a) Variation of the binding energy of $\Sigma_{QK0}$ with the threshold
$\omega_c$ and the Borel parameter $T$. From top to bottom the curves correspond to 
$\omega_c=1.8, 1.6, 1.4$ GeV. (b) The case for $\Sigma_{QK1}$. (c) For $\Sigma_{Qk1}$.
(d) For $\Lambda_{Qk1}$. }
\end{center}

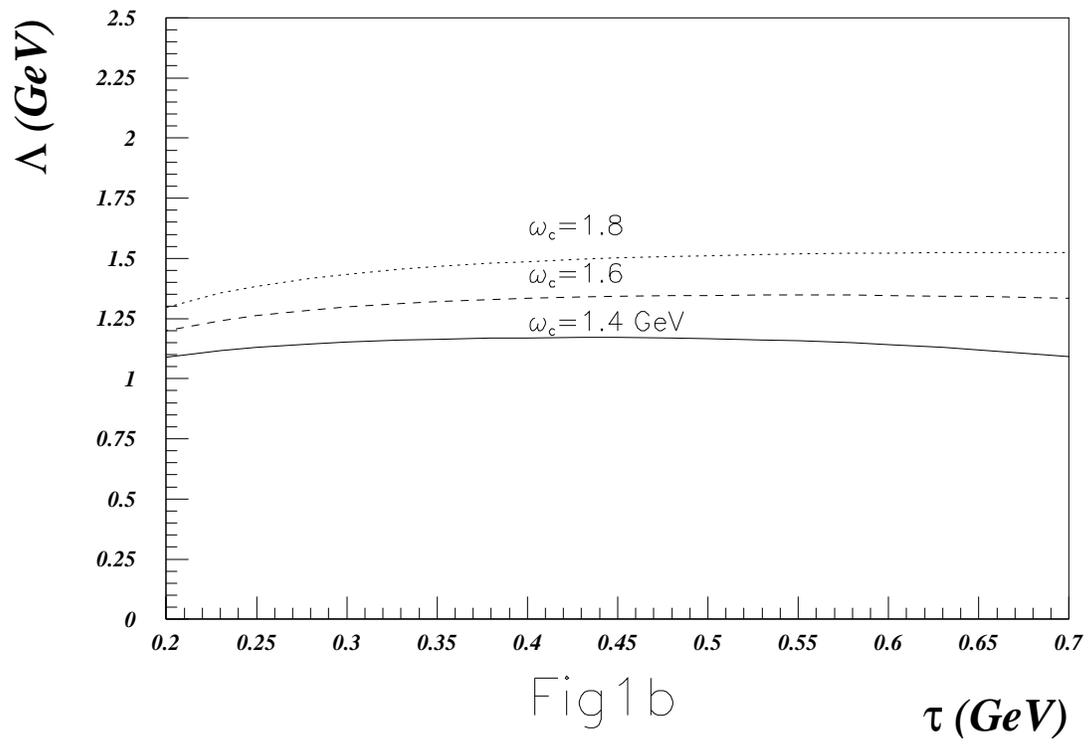
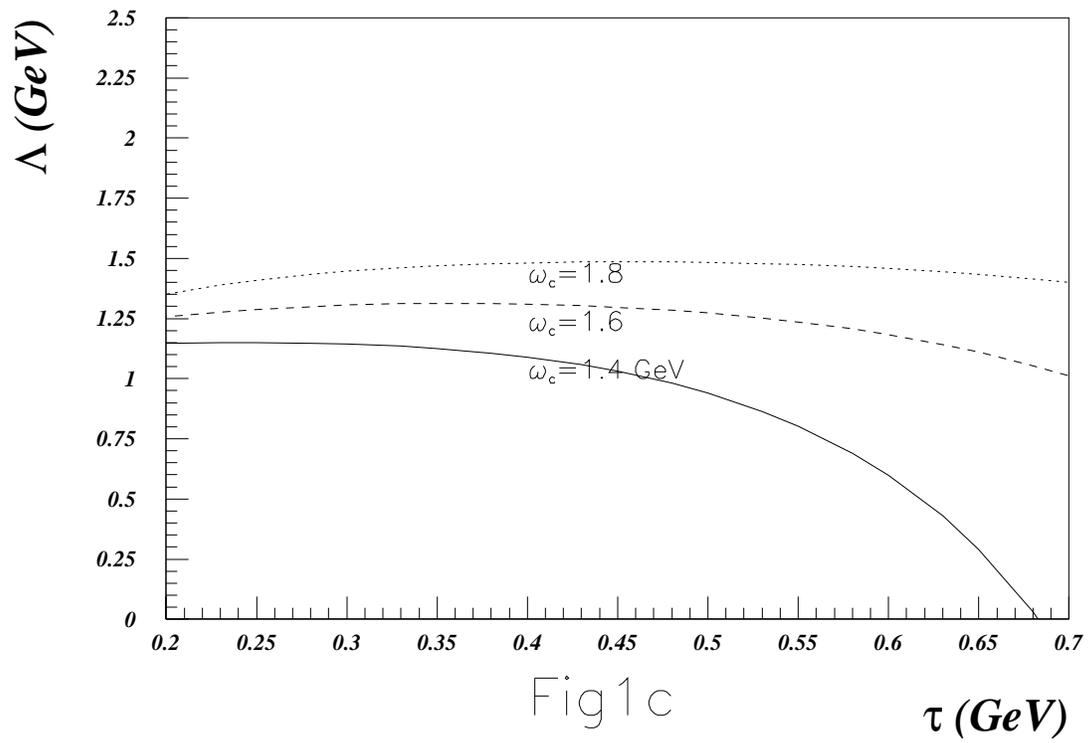
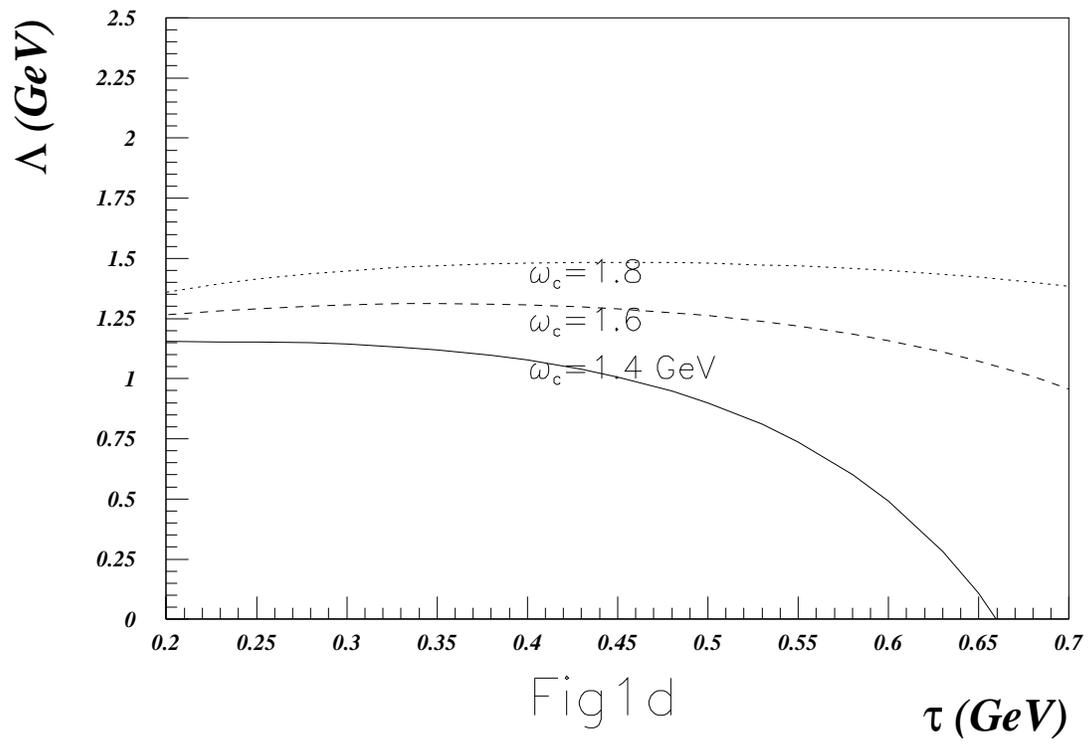
\end{document}